\documentclass[twocolumn,showpacs,preprintnumbers,amssymb,pre,superscriptaddress]{revtex4}

\usepackage{graphicx}
\usepackage{dcolumn}
\usepackage{bm}
\usepackage{amstext}

\begin{document}


\title{Vesicle shape, molecular tilt, and the suppression of necks}

\author{Hongyuan Jiang}
\affiliation{Division of Engineering, Box D, Brown University, Providence, RI 02912}
\author{Greg Huber}
\affiliation{Center for Cell Analysis \& Modeling, and Department of Cell Biology, University of Connecticut Health Center, Farmington CT 06032; and Department of Mathematics, University of Connecticut, Storrs, CT 06269}
\author{Robert A. Pelcovits}
\affiliation{Department of Physics, Brown University, Providence, RI 02912}
\author{Thomas R. Powers}
\affiliation{Division of Engineering, Box D, Brown University, Providence, RI 02912}
\email{Thomas_Powers@brown.edu}

\date{\today}


\begin{abstract}

Can the presence of molecular-tilt order significantly affect the shapes of lipid bilayer membranes, particularly membrane shapes with narrow necks?
Motivated by the propensity for tilt order and the common occurrence of narrow necks in the intermediate stages of biological processes such as endocytosis and vesicle trafficking, we examine how tilt order inhibits the formation of necks in the equilibrium shapes of vesicles.  For vesicles with a spherical topology, point defects in the molecular order with a total strength of $+2$ are required. We study axisymmetric shapes and suppose that there is a unit-strength defect at each pole of the vesicle. The model is further simplified by the assumption of tilt isotropy: invariance of the energy with respect to rotations of the molecules about the local membrane normal.  This isotropy condition leads to a minimal coupling of tilt order and curvature, giving a high energetic cost to regions with Gaussian curvature and tilt order.  Minimizing the elastic free energy with constraints of fixed area and fixed enclosed volume determines the allowed shapes. Using numerical calculations, we find several branches of solutions and identify them with the branches previously known for fluid membranes.  We find that tilt order changes the relative energy of the branches, suppressing thin necks by making them costly, leading to elongated prolate vesicles as a generic family of tilt-ordered membrane shapes.


\end{abstract}

\pacs{87.16.Dg, 61.30.Gd, 02.40.Hw, 46.70.Hg}



\maketitle

\section{Introduction}\label{Introduction}
The interplay of surface curvature and liquid-crystalline order finds its fullest expression in the manifold and complex biological realizations of the bilayer membranes surrounding cells and intracellular organelles. Helfrich~\cite{helfrich1973} was one of the first to connect membrane shape and molecular order, by realizing that
the spontaneous curvature of a bilayer membrane could arise from the spontaneous splay of the ordered rod-like lipid molecules comprising
the membrane.  But the biological world offers richer varieties of orientational order and shape that remain to be understood. For example, lipid molecules typically tilt relative to the normal of the membrane~\cite{Smith_etal1988,Smith_etal1990,Mouritsen2005}, and it has recently become possible to image tilt-ordered domains on the surface of
curved, micron-scale membranes~\cite{ZhaoMahajanLuFang2005}.
Furthermore, curvature has proven to play an active role in cellular and subcellular processes~\cite{McMahonGallop2005}.
Narrow necks with small mean curvature but large negative Gaussian curvature are relevant to biological membranes that
compartmentalize through budding, since this neck geometry allows separate membrane-bound compartments to be budded off, while avoiding high-energy membrane shapes.  Neck formation is universal and crucial to the phenomena of endo- and exocytosis~\cite{Frolov_etal2003,BaumgartHessWebb2003}, viral entry and budding, the traffic of continual fusion and fission of vesicle and Golgi membrane, and the interconnections between Golgi stacks~\cite{DergancMironovSvetina2006} and between the smooth and rough endoplasmic reticulum.  Because of the close association of these phenomena with cell function, it is crucial to understand the forces on membrane necks and the constraints on their formation.

\begin{figure} [htb]
    \centering
        \includegraphics[height=2.5in]{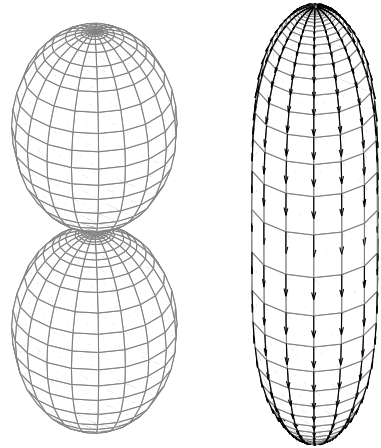}
        \caption{Effect of tilt order on membrane shape. Left: fluid membrane with reduced volume $v=0.706$ and spontaneous curvature $c_0=2.4/R_0$, where $R_0$ is an overall length scale defined in Sec.~\ref{modelanalysis}.
        Right: Tilt-ordered membrane with same parameters and tilt modulus $K_m=2.0\kappa$, where $\kappa$ is the bending stiffness of the membrane.  The arrows represent the tilt order.  A $+1$ defect sits at both the north  and the south poles of the vesicle.}
\label{twoshapes}
\end{figure}

In this paper we numerically calculate the equilibrium shapes of axisymmetric vesicles with tilt order. Figure~\ref{twoshapes} shows how dramatic the effect of tilt can be. The  two vesicles have the same resistance to bending, the same enclosed volume, and the same area, but the one on the right has tilt order whereas the one on the left has no tilt order. As we review below, the tilt order may be described by a vector field which is tangent to the vesicle surface. Molecular interactions prefer uniform tilt order, which may be realized on a surface with zero Gaussian curvature such as a plane or the surface of a cylinder. But it is impossible to have a uniform vector field on a surface with nonzero Gaussian curvature, such as a sphere or the neck connecting the two spheres. Therefore, uniform molecular order and Gaussian curvature are incompatible~\cite{MacKintoshLubensky1991,LubenskyProst1992}. In particular, as long as the molecular interactions are strong enough, the elongated prolate shape of Fig.~\ref{twoshapes} will be preferred over a shape with a neck.

We begin with a brief discussion of the relation of our work to previous work on membrane shapes and orientational order.
Section~\ref{modelanalysis} describes our minimal isotropic tilt model, coordinates, and numerical method. Our analysis and methods are straightforward, but we describe them here to make our paper self-contained. In Sec.~\ref{results} we present the main results, which are the energy as a function of reduced volume for several different branches of solutions, and  phase diagrams for shapes. In the final section we discuss the implications and limitations of our analysis.

\section{Relation to previous work}

The equilibrium shapes of closed fluid membrane vesicles have been studied theoretically and experimentally for many years (see~\cite{seifert1997,Dobereiner_etal1997}, and references therein). In the spontaneous-curvature model pioneered by Helfrich, a patch of membrane has a resistance to bending but is curved
in the absence of external loads~\cite{helfrich1973}. A different approach known as the bilayer-coupling model accounts for the bilayer structure of a membrane by imposing a constraint on the number of molecules in both monolayers~\cite{SvetinaZeks1983,SvetinaZeks1989}.  Both models predict the same set of vesicle shapes. However, the spontaneous-curvature model predicts that most shape transitions are discontinuous, while the bilayer-coupling model predicts continuous transitions~\cite{SeifertBerndlLipowsky1991}. To simplify our discussion, we will consider the spontaneous-curvature model with tilt order.

As alluded to above, the basic physics governing the interaction of vesicle shape and orientational order is the incompatibility of Gaussian curvature and uniform order. This incompatibility is a local property: a patch of surface with Gaussian curvature cannot have uniform tilt order.  The global topology of surfaces also constrains the number and strength of point defects in the orientation order field, via the Poinc\'are-Brouwer theorem, which states that the total defect strength of a vector field on a surface is equal to the Euler characteristic (see~\cite{kamien2002} for an elementary proof).  In our problem, a point defect is an isolated point where the tilt order vanishes, and the strength of the defect is the number of rotations of the tilt order field around that point.   For a vesicle with the topology of a sphere, the total defect strength is $+2$. It is natural to suppose that the lowest energy states have two $+1$ defects at antipodal points (we shall impose this two-defect configuration).

MacKintosh and Lubensky modeled a vesicle with spherical topology made up of molecules undergoing a transition from an untilted smectic-{\textit A} phase to a tilted smectic-{\textit C} phase~\cite{MacKintoshLubensky1991}.
They found that an initially spherical vesicle elongates into a prolate shape, with most of the Gaussian curvature concentrated near the defects that form at the two poles. They calculated the change in shape for this transition assuming fixed area, but they did not constrain the enclosed volume. Other work has examined the transitions among spherical, cylindrical, and toroidal vesicles with orientational order, again without the constraint of fixed enclosed volume~\cite{LubenskyProst1992,evans1995}.   In the current work, we impose the more realistic double constraint of fixed volume and fixed area, and solve for the shape. Also, our numerical method allows us to study shapes with large deflections from the spherical geometry. Therefore we can study the effect of tilt order on non-spherical shapes such as the pears and oblates predicted by the fluid membrane model. Rather than studying the transition in the tilt order (as in ~\cite{MacKintoshLubensky1991}), we focus on the shape effect: the effect of the tilt modulus $K_m$ (the elastic constant governing the resistance to non-uniform tilt order) on the overall membrane shape.

Topological defects can also form for geometrical reasons, even on surfaces such as tori which do not require any defects in orientational order. Our work is complementary to recent work on the formation and interaction of such defects on a fixed but arbitrarily curved surface~\cite{BowickNelsonTravesset2000}~\cite{VitelliNelson2004}. Instead of prescribing the shape and solving for the orientational order field, we prescribe the positions of two defects and solve for the vesicle shape and tilt field. We disallow additional defect formation, and discuss the validity and limitations of this restriction in Sec.~\ref{discuss}.

\section{The Model and Its Analysis}
\label{modelanalysis}

We make several simplifying assumptions in our analysis. Since the
vesicles we consider are much larger than their constituent
molecules, we use continuum mechanics in the long-wavelength
approximation.  Thermal fluctuations are disregarded.
We assume
that the vesicle shapes are surfaces of revolution, and the tilt
configuration is axisymmetric. In particular, the defects required
by topology are assumed to sit at the two poles of the vesicle. These
assumptions reduce the partial differential equations governing
the shape and tilt configuration to ordinary differential
equations, which greatly simplifies our calculations.  Also, we
suppose that the bilayer membrane is thin compared to the
characteristic size of the vesicle. Since the lipid bilayers are
approximately two nanometers thick~\cite{Mouritsen2005}, this
assumption is highly accurate for vesicles of micron size and
larger. A consequence of this assumption is that stretching is
much more costly than bending; therefore, we demand that the total
area remain constant. And, although membranes are permeable to
water, osmotic effects resist changes in
volume~\cite{seifert1997}, leading us to fix the volume. As we
explain below, we use a minimal model for the orientational order,
disregarding anisotropic couplings between the tilt field and the
membrane curvature.
We also disregard chiral interactions, which
have been shown to be important in models for lipid tubule
structure~\cite{selinger_mackintosh_schnur1996}~\cite{TuSeifert2007}
and a proposed mechanism for budding~\cite{SarasijRao2002,SarasijMayorRao2007}.

We will study how vesicle shape depends on area and volume in the presence of tilt order. Just as in the case of fluid membranes, we will see that the bending energy and tilt stiffness energy terms are scale invariant. This invariance allows us to vary area $A$ and volume $V$  by changing one parameter, the reduced volume $v$~\cite{seifert1997}. The reduced volume is the ratio of the actual volume of a vesicle to the volume of a sphere with the same area as the vesicle. If $R_0$ is the radius of the sphere with area $A$, then $v=V/(4\pi R_0^3/3)$.
\subsection{Parametrization and geometry}

\begin{figure} [htb]
    \centering
        \includegraphics[height=2.5in]{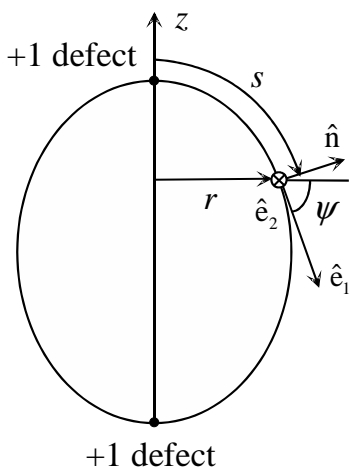}
        \caption{Vesicle coordinate system and assumed location of defects.
        The shape is a surface of revolution about the $z$ axis.
        The vector $\mathbf{\hat e}_2$ points into the page as indicated by the $\bigotimes$. }
\label{setup}
\end{figure}

In this section we describe our parametrization and fix the notation; see~\cite{David1989} and~\cite{kamien2002} for general discussions of differential geometry applied to membranes.
We choose the $z$ axis to be the axis of symmetry of the vesicle and represent points on the surface of the vesicle by the three--dimensional vector $\mathbf{X}(\phi,s)=(r(s) \cos \phi,r(s)\sin \phi,z(s))$, where $\phi$ and $r$
are plane polar coordinates in the $xy$ plane, and $s$ is the arclength measured from the north pole of the surface along a line of longitude (see Fig.~\ref{setup}).  Define $\psi(s)$ to be the angle between the tangent vector $\partial_s \mathbf{X}$ along a longitude and the horizontal axis.   Then $\mathrm{d}z/\mathrm{d}r=z_s/r_s=-\tan \psi(s)$, where $0<\psi(s)<\pi$. In terms of $\psi$, we have $r_s=\cos\psi$ and  $z_s=-\sin \psi$.

An orthonormal basis in the tangent plane of the surface is given by
\begin{eqnarray}
\nonumber\mathbf{\hat e}_1 &=& \partial_s \mathbf{X}
= (\cos \psi \cos\phi,\cos \psi \sin\phi, -\sin \psi)\\
\mathbf{\hat e}_2 &=&  \partial_\phi\mathbf{X} /| \partial_\phi\mathbf{X} |
= (-\sin\phi,\cos\phi,0)\label{e1},
\label{e2}
\end{eqnarray}
where the $s$ and $\phi$  subscripts denote partial differentiation with respect to the coordinates $s$ and $\phi$, respectively.
We construct the outward normal $\mathbf{\hat n}$ to the surface using the orthonormal frame:
\begin{equation}
\mathbf{\hat n} = \mathbf{\hat e}_1 \times
\mathbf{\hat e}_2=(\sin\psi\cos\phi,\sin\psi\sin \phi,\cos\psi)\label{n}.
\end{equation}

 The metric tensor $g_{ij}$ of the surface is given by
\begin{eqnarray}
 g_{ij}&=&\partial_i \mathbf{X}\cdot \partial_j \mathbf{X} = \left( \begin{array}{cc}
1 & 0 \\
0 & r^2\end{array} \right)\label{g},
\end{eqnarray}
where the indices $i,j=1,2$ label the coordinates $s$ and $\phi$, respectively. As usual, we denote the inverse of the metric tensor by $g^{ij}$, and we use $g^{ij}$ to raise indices. The second fundamental form $K_{ij}$ is defined by
\begin{equation}
K_{ij} \equiv \mathbf{\hat n}\cdot\partial_i \partial_j\mathbf{X}=-\left( \begin{array}{cc}
\psi_s & 0 \\
0 & r\sin\psi\end{array} \right).
\end{equation}
From the second fundamental form we construct the mean curvature $H$ and Gaussian curvature $K$:
\begin{eqnarray}
H &\equiv&-\frac{1}{2} g^{ij}K_{ij}=\frac{1}{2}(\psi_s+\sin\psi/r)\label{H}\\
K &\equiv&\det({K^i}_j)=\det(g^{il}K_{lj})=\psi_s \sin\psi/r,
\end{eqnarray}
where repeated indices have been summed over. Note from Eq.~(\ref{H}) that we use a convention in which the sphere has positive mean curvature.

We assume that the lipid molecules on the surface of the vesicle are tilted at a preferred angle with respect to the surface normal $\mathbf{\hat n}$, as in a smectic-\textit{C} phase~\cite{deGennesProst}. Let the vector $\mathbf m$ denote the projection of the directors of the lipid molecules onto the tangent plane. Since we do not study the transition between tilted and untilted phases, it is convenient to normalize $\mathbf m$ to make $|\mathbf m|=1$ well away from topological defects.   In terms of the local orthonormal basis of the tangent plane, we have
\begin{equation}
\mathbf{m}=B(\cos{\theta}\mathbf{\hat e}_1+\sin{\theta}\mathbf{\hat e}_2)=B\mathbf{\hat{m}},
\label{B}
\end{equation}
where the amplitude $B$ vanishes at defect centers, and approaches unity far from defect cores.

We must assign an energy penalty to nonuniform configurations of the tilt field $\mathbf m$. For a flat surface, $\mathbf m$ is uniform if the components of $\mathbf m$ are constant in the standard Cartesian basis. Thus, a suitable energy density would be proportional to $\partial_i m^j\partial_i m^j$.
However, on a curved surface, $\mathbf m$ can vary with position not only because $m^i$ varies with position, but also because $\mathbf{\hat e}_i$ can vary. Furthermore, it is only the tangential component of derivatives of $\mathbf m$ that enter the tilt stiffness terms; normal components add to the resistance to bending and therefore may be absorbed in the bending energy term, discussed below. These features are captured by the covariant derivative
\begin{eqnarray}
D_i\mathbf{m}&=&\partial_i\mathbf{m}-(\mathbf{\hat n}\cdot\partial_i\mathbf{m})\mathbf{\hat n}\nonumber\\
&=&(\partial_i m^1-m^2 \Omega_i) \mathbf{\hat e}_1+(\partial_i
m^2+m^1\Omega_i) \mathbf{\hat e}_2,\label{Dm}
\end{eqnarray}
where the ``spin connection,"
\begin{equation}
\Omega_i=\mathbf{\hat e}_2\cdot\partial_i \mathbf{\hat e}_1,\label{Omega1}
\end{equation}
 is the rate at which the frame
$\{\mathbf{\hat e}_1,\mathbf{\hat e}_2\}$ rotates about the normal $\mathbf{\hat n}$ as the $i$th coordinate increases. The covariant curl of the spin connection is the Gaussian curvature:
\begin{equation}
K=-\frac{1}{\sqrt{g}}\epsilon_{ij}\partial_i\Omega_j,\label{curlOmega}
\end{equation}
where $g$ is the determinant of the metric tensor and
$\epsilon_{ij}$ is the antisymmetric symbol with
$\epsilon_{12}=1$~\cite{kamien2002}. We can now see why Gaussian
curvature is incompatible with a uniform tilt field. Consider a
tilt field on a surface of nonzero Gaussian curvature. A tilt
field on a curved surface is uniform if it has a vanishing
covariant derivative. Writing Eq.~(\ref{Dm}) in terms of $B$ and
$\theta$, we find that $D_i{\mathbf m}=0$ implies $\partial_i B=0$
and $\partial_i\theta+\Omega_i=0$. Consider a patch on a curved
surface that includes no defects, so that $\theta$ is smooth. Then
to solve $\partial_i\theta+\Omega_i=0$ for $\theta$ we must have
$\epsilon_{ij}\partial_i\Omega_j=0$~\cite{kamien2002}. Therefore,
the tilt field cannot be uniform on a patch with Gaussian
curvature.

For our parametrization,
$\Omega_i=(0,\cos\psi)$: the frame $\{\mathbf{\hat e}_1,\mathbf{\hat e}_2\}$ rotates about the normal as $\phi$ changes, but not as $s$ changes. Consistent with our assumption that the shape of the surface is axisymmetric, we assume that the orientation of the molecules is axisymmetric as well, with $\partial_\phi B=0$ and $\partial_\phi \theta=0$. Therefore,
\begin{eqnarray}
D_\phi\mathbf{m}=B\cos\psi\mathbf{\hat{m}}_{\perp}\label{Dphi}\\
D_s\mathbf{m}=B_s\mathbf{\hat{m}}+B\theta_s\mathbf{\hat{m}}_{\perp}\label{Ds},
\end{eqnarray}
where $\mathbf{\hat{m}}_{\perp}=-\sin\theta\mathbf{\hat e}_1+\cos\theta\mathbf{\hat e}_2$, $B_s=\partial_s B$, and $\theta_s=\partial_s\theta$.

\subsection{Free energy}

The free energy $F$ of the vesicle is the sum of terms associated with the bending of the vesicle and terms associated with the tilt vector order parameter field: $F=F_b+F_m$. We use the Helfrich model for bending energy,
\begin{equation}
\label{Fb}F_b =\int \left[\frac{\kappa}{2}(2H-c_0)^2+\kappa_G K\right]\sqrt{g}\,\mathrm{d}s \mathrm{d}\phi,
\end{equation}
In Eq.~(\ref{Fb}), $\kappa$ is the bending modulus, typically $10$--$15k_\mathrm{B}T$~\cite{Mouritsen2005}, and $\kappa_G$ is the Gaussian rigidity.
The spontaneous curvature $c_0$ is twice the preferred value of the mean curvature for a patch of membrane.
Spontaneous curvature of a bilayer membrane can arise either from the sum of the inherent spontaneous curvatures of the monolayers,
or from a difference in the number of molecules in either monolayer~\cite{DobereinerSelchowLipowsky1999}. Since we consider a closed surface with fixed topology, the Gauss-Bonnet theorem ensures that the integral of the Gaussian curvature is independent of shape and contributes only an overall constant to the free energy~\cite{MillmanParker1977}. Therefore the term proportional to $\kappa_G$ may safely be disregarded.

The elastic free energy $F_m$ for the tilt order is a sum of many terms, including costs for splay and bend of the director field, and many terms coupling the director field to the vesicle shape~\cite{nelson_powers1993,powers_nelson1995}. To simplify our task, we demand that the energy be isotropic in tilt, i.e.~invariant under arbitrary rotations of $\mathbf{m}$  about the normal $\mathbf{\hat n}$. This symmetry rules out all of the anisotropic terms, leaving only a minimal coupling of the tilt order to shape:
\begin{eqnarray}
\label{Fm}F_m &=& \frac{1}{2}\int\Bigl[ K_m D_i m^j D^i m_j\nonumber \\
& &+\frac{\lambda}{2} (1-m_i m^i)^2\Bigr]\sqrt{g}\,\mathrm{d}s \mathrm{d}\phi.
\end{eqnarray}
The first term of Eq.~(\ref{Fm}) gives a preference for a uniform tilt field. Since we impose isotropy, the free energy density at a point is independent of the direction of $\mathbf{m}$ relative to the principal directions of curvature. (We discuss how anisotropy may affect our results in Sec~\ref{discuss}). Note that the Frank elastic constant $K_m$ has the same dimensions as $\kappa$, implying that the effects of this term are comparable to the effects of the bending term of Eq.~(\ref{Fb}). The second term of Eq.~(\ref{Fm}) gives a preference for $|\mathbf{m}|=1$. We assume that we are deep in the ordered phase, so that $\lambda R_0^2/K_m\gg1$. The length scale $\sqrt{K_m/\lambda}$ determines the radius of the defect core, wherein $|\mathbf{m}|$ falls steeply to zero.

The shape of the vesicle and the orientation of the tilted molecules on its surface are determined by minimizing $F$ subject to a
given surface area $A$ and volume $V$. To impose these constraints, we introduce Lagrange multipliers $\Sigma$ and $P$. It is convenient to treat $r$ and $\psi$ as independent variables in the variation of the free energy. Therefore, we introduce an additional Lagrange multiplier function
$\gamma(s)$ to impose the local constraint $r_s=\cos\psi$. Thus, our task is to minimize
\begin{equation}
F^\prime\equiv F+\Sigma A+ P V+\kappa\int\gamma(s)(r_s-\cos\psi)\mathrm{d}s\mathrm{d}\phi.
\end{equation}
It is convenient to scale the Lagrange multipliers by $\kappa$: $\bar\Sigma=\Sigma/\kappa$ and $\bar P=P/\kappa$. Then $F^\prime$ can be written as
\begin{eqnarray}
F^\prime &=& 2\pi\kappa\int_0^{L} f^\prime(\psi,\psi_s,r,r_s,B,B_s,\theta_s,\gamma) \,\mathrm{d}s,
\label{energy2}
\end{eqnarray}
where the upper integration limit $L$ is the total arclength along a longitude from the north to the south pole, and
\begin{eqnarray}
&&f^\prime(\psi,\psi_s,r,r_s,B,B_s,\theta_s,\gamma)\nonumber\\
&& = r\Biggl\lbrack\frac{1}{2}(\sin\psi/r+\psi_s-c_0)^2+\bar\Sigma + \frac{1}{2}\bar P r\sin\psi \nonumber\\
&& + \frac{\lambda}{4\kappa}(1-B^2)^2 +\frac{K_m}{2\kappa}(B^2{\cos^2\psi}/r^2+{B_s}^2+B^2 \theta_s^2)\Biggr\rbrack\nonumber\\
&& + \gamma(s) (r_s-\cos\psi).\label{fp}
\end{eqnarray}

\subsection{Euler-Lagrange equations}

The Euler-Lagrange equations which extremize $F^\prime$ are given by
\begin{eqnarray}
\psi_s&=&\frac{U}{r}-\frac{\sin\psi}{r}+c_0,\label{el1}\\
U_s&=&\frac{U}{r}\cos\psi+\gamma \sin\psi
+\frac{1}{2}\bar P r^2\cos\psi\nonumber\\
&  & -\frac{K_m B^2}{\kappa r}\cos\psi \sin\psi,\\
B_s&=&\frac{\kappa W}{K_m r},\label{Bs}\\
W_s&=&-\frac{\lambda}{\kappa} rB(1-B^2)\nonumber\\
& &+\frac{K_m r B}{\kappa}\left(\frac{\cos^2\psi}{r^2}+\theta_s^2\right),\label{Ws}\\
\gamma_s&=&\frac{U^2}{2r^2}-\frac{U}{r^2}\sin\psi+\bar Pr\sin\psi+\bar\Sigma\nonumber\\
& &+\frac{K_m}{2\kappa}\left[-\frac{B^2}{r^2}\cos^2\psi+\left(\frac{\kappa W}{K_m r}+B^2\theta_s^2\right)^2\right]\nonumber\\
& &+\frac{\lambda}{4\kappa}(1-B^2)^2,\label{gammaseq}\\
r_s&=&\cos\psi,\label{rseq}\\
\left(r B^2 \theta_s \right)_s&=&0.\label{thetas}
\end{eqnarray}
We have introduced two auxiliary functions $U$ and $W$ to obtain first-order differential equations, which are required for our numerical routine.

The Euler-Lagrange equations can be simplified somewhat by the
observation that $\theta$ is independent of $s$.
Equation~(\ref{thetas}) implies that $r B^2 \theta_s$ is
independent of $s$. However, because $r$ and $B$ both vanish at
the poles, $r K_m^\prime B^2 \theta_s=0$ everywhere. Since $B$ and
$r$ are nonzero only right at the poles,  $\theta_s=0$ everywhere.
This result can be seen more directly by examining the form of the free
energy $F_m$, Eq.~(\ref{Fm}).  Since $\Omega_s=\Omega_1=0$, $F_m$
is minimized when $\theta$ is independent of $s$.
Note that the
isotropy of the energy means that $F$ is independent of the angle
$\theta$.  In Fig.~\ref{twoshapes}, for example, we chose to have
the directors aligned with lines of latitude.   The choice was arbitrary -- directors aligned
along lines of longitude, or any other direction, would result in the same free energy and same shape.

For the remaining six equations (\ref{el1}--\ref{rseq}) we have six corresponding boundary conditions. The angle $\psi$ and radius $r$ are fixed at either pole: $\psi(0)= r(0)=r(L)=0$, and $\psi(L)=\pi$. The amplitude $B$ at either pole vanishes due to the assumed presence of defects, $B(0)=B(L)=0$. However, $L$ is still unknown and must be solved for along with the shape. To determine $L$, we reparametrize the problem, introducing a new independent variable $t$ such that $t=0$ at the north pole and $t=1$ at the south pole: $s=L t$, where $L$ is a constant~\cite{JulicherLipowsky1996}.
The Euler-Lagrange equations~(\ref{el1}--\ref{thetas}) are therefore modified by replacing $\mathrm{d}/\mathrm{d}s$ with $(1/L)\mathrm{d}/\mathrm{d}t$. Note that the boundary conditions on $r$, $\psi$, and $B$ at $s=L$ become boundary conditions at $t=1$.
The additional equation allowing us to solve for $L$ is $\mathrm{d}L/\mathrm{d}t=0$. To determine the additional boundary condition required by this equation, we consider the symmetry of the free energy under a constant shift in $s$. The free energy density now takes the form  $F'=2\pi\kappa\int_0^1\tilde  f\mathrm{d}t$, where
\begin{eqnarray}
&&\tilde f(\psi,\psi_t,r,r_t,B,B_t,\theta_t,\gamma,s_t)\nonumber\\
&&=s_t f^\prime(\psi,\psi_s/L,r,r_s/s_t,B,B_s/s_t,\theta_s/s_t,\gamma),
\end{eqnarray}
and $s_t=L$.
Since $s(t)$ does not appear explicitly in the free energy density, there is a first integral or Hamiltonian function $\partial \tilde f/\partial s_t$ which is independent of $t$. Examination of the variation of $\tilde f$ with respect to $s(t)$ leads to $\partial \tilde f/\partial s_t=0$ at $t=1$, implying that the Hamiltonian function vanishes everywhere. Writing the Hamiltonian explicitly in terms of the dependent variables at $t=0$ yields the desired boundary condition $\gamma(0)=0$~\cite{SeifertBerndlLipowsky1991,JulicherLipowsky1996}.

To complete our specification of the shape equations, we describe how we implement the constraints of fixed area and volume. It is convenient to regard area as a function of $t$. Define $A(t)$ as the area of the portion of the vesicle surface north of the line of latitude corresponding to $t$, with a similar definition for $V(t)$. Then
\begin{eqnarray}
A_t&=&2\pi r L\label{At}\\
V_t&=&\pi r^2\sin\psi L\label{Vt}\\
\bar\Sigma_t&=&0\label{sigmat}\\
\bar P_t&=&0,\label{Pt}
\end{eqnarray}
where Eqs.~(\ref{sigmat})--(\ref{Pt}) arise because the Lagrange multipliers are constant, but unknown and shape-dependent.
The four boundary conditions corresponding to Eqs.~(\ref{At}--\ref{Pt}) are $A(0)=0$, $A(1)=4\pi R_0^2$, $V(0)=0$, and $V(1)=4\pi R_0^3v/3$. The scale $R_0$ is set to unity in our numerical calculations. Finally, the shape is determined by integrating $z_t=-L\sin\psi$ with boundary condition $z(0)=0$.
\begin{figure*}
    \centering
        \includegraphics[height=5.in]{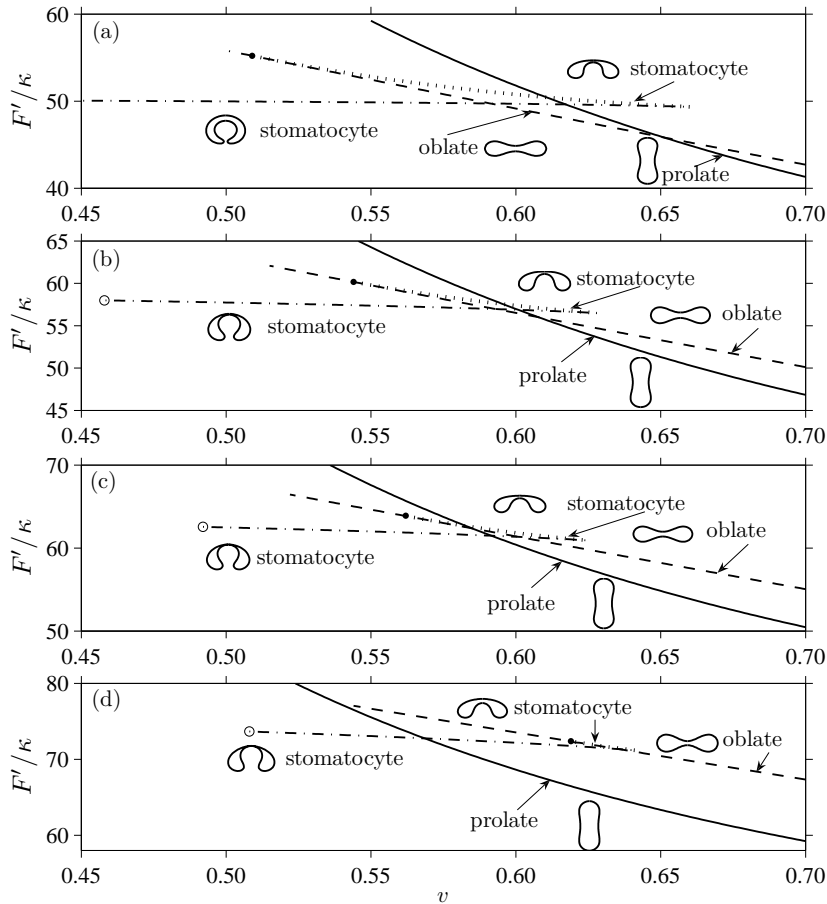}
        \caption{ Dimensionless free energy $F'/\kappa$ vs. reduced volume $v$ for fluid membrane vesicles with $c_0=0$ for three different values of Frank elastic constant:  (a)  $K_m=0$.  (b)  $K_m=0.3\kappa$.  (c) $K_m=0.5\kappa$ (d) $K_m=\kappa$. Solid lines correspond to prolate shapes, dashed lines correspond to oblate shapes, and the dash-dotted and dotted lines correspond to stomatocyte shapes. The filled circles denote continuous bifurcations. The open circles denote limit points at which the vesicle intersects itself.}
\label{c0_0_2}
\end{figure*}

\begin{figure} [htb]
    \centering
        \includegraphics[height=2.5in]{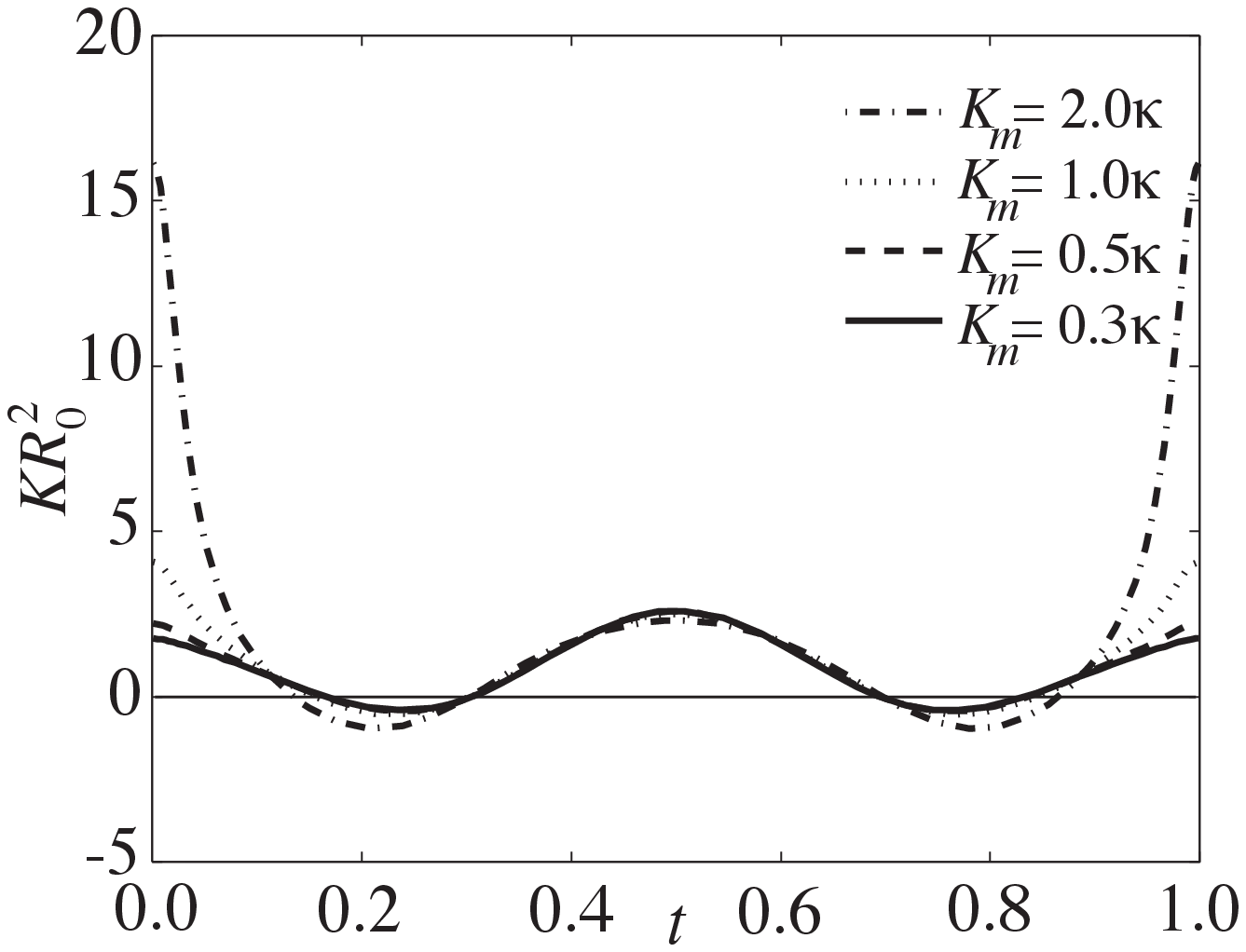}
        \caption{Dimensionless Gaussian curvature $KR_0^2$ versus the dimensionless parameter $t$ for oblate shapes with $c_0=0$ and various values of $K_m/\kappa$. Note that $t=0$ at the north pole, $t=1$ at the south pole, and $t=0.5$ at the equator, which lies in the horizontal plane midway between the two
defects. }
\label{GKplot}
\end{figure}

Although the Euler-Lagrange equations are nonlinear and must be solved numerically, it is straightforward to determine the form of the amplitude $B$ at the centers of the defect cores. For example, near $s=0$, we have $\cos\psi\approx0$ and $r\approx s$. Therefore, equations~(\ref{Bs}) and (\ref{Ws}) reduce to Bessel's equation for $s\approx0$:
\begin{eqnarray}
s^2 B_{ss}+s B_s+ \left(\frac{\lambda }{K_m} s^2-1\right)B=0,\label{bessel}
\end{eqnarray}
and thus $B\propto s$ in the core, which has a size set by $\sqrt{K_m/\lambda}$, as mentioned earlier. Similar considerations apply to the defect located at the south pole.

In our numerical approach, we treat the Euler--Lagrange  equations as a two-point boundary value problem, using the MATLAB function {\tt bvp4c}~\cite{ShampineGladwellThompson2003}.
We choose $\lambda/K_m= 1000$ to make the defect cores small.
To avoid the divergences that occur in the equations when
$r\rightarrow0$, we solve the equations in the interval
$\delta<t<1-\delta$, where $\delta =0.001$. All boundary conditions are now evaluated at $t=\delta$ or $t=1-\delta$. We use Taylor series to relate the boundary values at these new endpoints to the values at $t=0$ and $t=1$:
\begin{eqnarray}
\psi(\delta)&\approx& \psi(0)+\psi_t(0)\delta=\psi_t(0)\delta\\
B(\delta)&\approx& B(0)+B_t(0)\delta=B_t(0)\delta\\
\psi(1-\delta)&\approx&\psi(1)-\psi_t(1)\delta=\pi-\psi_t(1)\delta\\
B(1-\delta)&\approx&B(1)-B_t(1)\delta=-B_t(1)\delta.\label{}
\end{eqnarray}
The constants $\psi_t(0)$, $B_t(0)$,
$\psi_t(1)$, and $B_t(1)$  are unknown parameters that may be determined
by the numerical routine using the conditions
\begin{eqnarray}
U(\delta) &\approx& L \psi_t(\delta) r(\delta)+\sin(\psi(\delta))+c_0 r(\delta)\\
U(1-\delta) &\approx& L \psi_t(1-\delta) r(1-\delta)\nonumber\\
&+&\sin(\psi(1-\delta))+c_0 r(1-\delta)\\
W(\delta) &\approx& L B_t(\delta) K_m r(\delta)/\kappa\\
W(1-\delta) &\approx& L B_t(1-\delta) K_m r(1-\delta)/\kappa.\label{}
\end{eqnarray}
from Eqns.~(\ref{el1}) and (\ref{Bs}). We also have
\begin{eqnarray}
\gamma(\delta)&=&\gamma(0)+\gamma_t(0)\delta\nonumber\\
 &\approx&\gamma(0)+\gamma_t(\delta)\delta=\gamma_t(\delta)\delta, \label{gdelta}
\end{eqnarray}
with an error of order $\delta^2$. The parameter $\gamma_t(\delta)$ is given in terms of the variables at $t=\delta$ by Eqn.~(\ref{gammaseq}).

\section{Results}\label{results}

For purposes of comparison with the case of a fluid membrane with no tilt order, we have
repeated the calculations of Ref.~\cite{SeifertBerndlLipowsky1991}
for  spontaneous curvature $c_0=0$ and $c_0R_0=2.4$. At $v=1$, the
vesicle shape is always spherical, with bending energy
$8\pi\kappa$. First consider the case $c_0=0$.  As $v$ is
decreased from unity, the lowest energy shape becomes prolate,
elongating continuously. Upon further decrease of $v$ the shape
changes discontinuously to oblate, and finally to the stomatocyte
shape. At $v=0$ where the dash-dotted line in Figure~\ref{c0_0_2}a
terminates, the stomatocyte shape consists of two concentric spheres, spaced infinitesimally
close, connected by a vanishingly thin
neck of zero mean curvature, with a total bending energy
$16\pi\kappa$. Figure~\ref{c0_0_2}a shows some of these shapes,
and the free energy vs. $v$ over the range for which the
transitions occur. Note that there are two stomatocyte branches;
the upper one (dotted line) is metastable.  As the reduced volume
is decreased along the upper stomatocyte  branch, the vesicle
height along the $z$ axis decreases and the shape becomes more
symmetric about the horizontal plane midway between the two
defects. The two branches join at the filled circle, where the
stomatocyte shape becomes oblate.  Figures~\ref{c0_0_2}b and c
show the effect of tilt order. As $K_m/\kappa$ increases, the
prolate branch of solutions has lower free energy than the oblate
branch for a greater range of reduced volume, until eventually the
oblate branch becomes completely metastable. To understand why
tilt order favors prolate shapes over oblate, note that as $v$
decreases, the prolate shapes extend more along the $z$ axis and
become more cylindrical, leading to a greater region of small
Gaussian curvature and approximately uniform tilt order. The oblate
shapes also have regions of small Gaussian curvature, but these
regions are confined to narrow bands near $t\approx0.2$ and $t\approx0.8$ (Fig.~\ref{GKplot}).

Figure~\ref{GKplot} also shows that our assumption that the defects are constrained to lie at the poles of the vesicle is reasonable.
Defects of positive sign prefer regions of positive
Gaussian curvature~\cite{BowickNelsonTravesset2004}. Figure~\ref{GKplot} shows that for $K_m/\kappa\lesssim0.5$, the Gaussian curvature at the equator is just greater than the Gaussian curvature at either pole. Thus, for $K_m\lesssim0.5\kappa$, we expect that the configuration with two defects at antipodal points on the equator might have smaller energy than the configuration with defects on either pole. Our
simplifying assumption of axisymmetry prevents us from
investigating this possibility. However, as $K_m/\kappa$ is increased to unity, the regions of maximum Gaussian curvature lie at the poles, and we expect that the minimum energy configuration is to have the defects lie at the poles. For the prolate shapes and the pear shapes considered below, the poles are always regions of maximum Gaussian curvature. Even for the stomatocyte, the north pole is the region of maximum Gaussian curvature, and the Gaussian curvature is roughly constant over much of the inner surface of the pocket.

\begin{figure} [htb]
    \centering
        \includegraphics[height=2.5in]{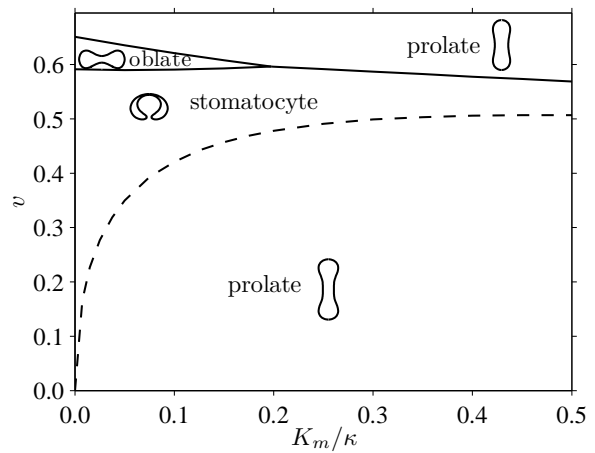}
        \caption{Phase diagram for lowest energy vesicle shapes for $c_0=0$.  The lowest energy shapes are shown for each value of $v$ and $K_m/\kappa$. The dashed line denotes the line of limit points for the stomatocyte shapes.}
\label{c00mu}
\end{figure}

A second qualitative effect of tilt order on the $c_0=0$ shapes is
that the lower stomatocyte branch develops a limit point
corresponding to self-intersection at the sites of the defects.
These limit points are denoted by open circles in
Fig.~\ref{c0_0_2}b and c. Figure~\ref{c00mu} shows the phase diagram for the lowest energy
vesicle shapes for the case $c_0=0$ as a function of the Frank
constant. The dashed line shows the values of $K_m/\kappa$ and $v$
for which the stomatocyte shape intersects itself. For values of
$v$ below the dashed line, the prolate branch again becomes the
lowest energy branch of solutions.

\begin{figure*} [htb]
    \centering
        \includegraphics[height=5.in]{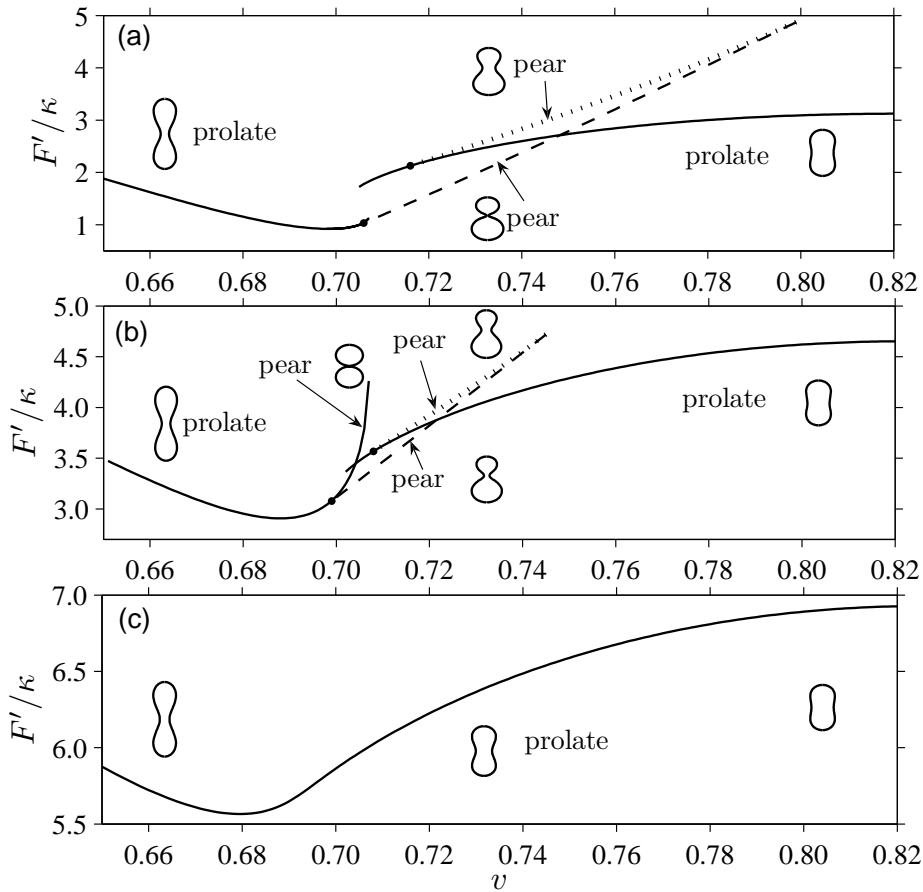}
        \caption{Dimensionless free energy $F'/\kappa$ vs. reduced volume $v$  for fluid membrane vesicles with $c_0R_0=2.4$ and three different values of Frank elastic constant: (a) $K_m/\kappa=0$,  (b) $K_m/\kappa=0.08$, (c) $K_m/\kappa=2.0$. Solid lines correspond to prolate shapes, dashed lines correspond to asymmetric pear shapes, and the dotted line corresponds to symmetric pear shapes. The filled circles mark continuous bifurcations. There is a symmetric pear branch in (a), just to the right of the first bifurcation, but it is too short to be seen. }
\label{c0_24_2}
\end{figure*}

\begin{figure} [htb]
    \centering
        \includegraphics[height=2.5in]{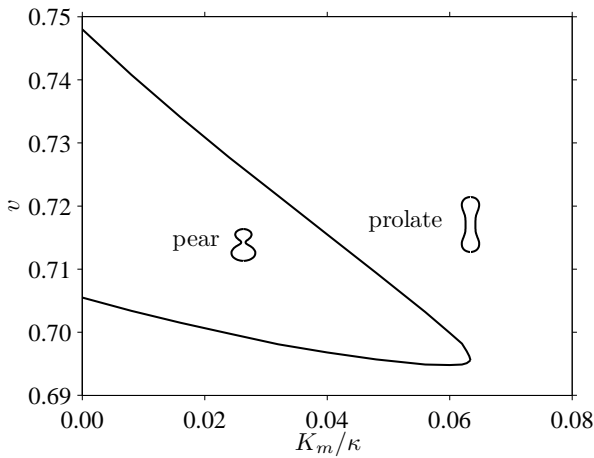}
        \caption{Phase diagram for $c_0R_0=2.4$.  The lowest energy shapes are shown for each value of $v$ and $K_m/\kappa$. }
\label{c024mu}
\end{figure}

Turning to the case $c_0R_0=2.4$, we see from
Ref.~\cite{SeifertBerndlLipowsky1991} that the fluid membrane
shapes are dominated by prolate shapes and pear shapes,
Fig.~\ref{c0_24_2}a. Note that the energy scale on the vertical
axis is much less than that of Fig.~\ref{c0_0_2}a since the mean
curvature required by the constraints of fixed volume and area is
close to $c_0$. The solid dots again denote continuous transitions
between pear and prolate shapes. The effect of tilt order is to
increase the energy cost of the pear shapes relative to the
prolate shapes, eventually making the prolate branch of solutions
the lowest energy branch for all $v$. It is interesting to note
that tilt order does not completely rule out pear shapes with
narrow necks. The phase diagram of lowest energy shapes in
Fig.~\ref{c024mu} shows that pear shapes are allowed in a range of
$v$ for sufficiently small $K_m$. The neck of the pear shape
becomes wider as $K_m$ increases. For these shapes, there is a
slight reduction in the order $B$ in a narrow band around the
neck.

\section{Discussion and conclusion}\label{discuss}

Our detailed, systematic calculations show that tilt order suppresses necks and favors elongated prolate shapes for sufficiently large tilt modulus $K_m$. We calculated the free energy as a function of reduced volume for several branches of solutions, and showed how increasing the tilt modulus increases the energy of the non-prolate branches relative to the prolate branches,
finally leading to a single family of prolate shapes.

Our calculation was based on several important assumptions. We ruled out many of the possible terms in the free energy by taking the free energy to be invariant under arbitrary rotations of $\mathbf{\hat m}$ about the normal $\mathbf{\hat n}$.   A natural way to remove this assumption and study the effects of anisotropy without introducing an unmanageable number of terms would be to replace $F_m$ with $F_d$, the one-Frank-constant approximation for the orientational free energy of the directors $\mathbf{\hat d}$ making up the membrane surface~\cite{selinger_mackintosh_schnur1996}:
\begin{equation}
F_d=\frac{K_m}{2}\int \left(\partial_i\mathbf{\hat d}\cdot\partial_j\mathbf{\hat d}\right)g^{ij}\sqrt{g}\mathrm{d}s\mathrm{d}\phi.\label{3diso}
\end{equation}
To compare $F_d$ with $F_m$, write $\mathbf{\hat d}=\alpha \mathbf{\hat N}+\mathbf{m}$, where $\mathbf{\hat N}$ is the local unit surface normal, and $\alpha$ is determined by $|\mathbf{\hat d}|=1$. Rewriting $F_d$ in terms of intrinsically two-dimensional quantities, we find that
\begin{eqnarray}
F_d&=&\frac{K_m}{2}\int \left[ g^{ij}\nabla_im^k\nabla_jm_k+\alpha^2K_i^jK_j^i \right.\nonumber\\
&+&\left. m^iK_{ij}K^j_km^k-2\alpha\left(\nabla_i m^j\right)K^i_j\right]\sqrt{g}\mathrm{d}s\mathrm{d}\phi.\label{threedtwod}
\end{eqnarray}
The first two terms of the integrand in (\ref{threedtwod}) are isotropic.
To see that the term $K_i^jK_j^i$ is already accounted for in our minimal isotropic tilt model, note that a matrix satisfies its own characteristic equation:
\begin{equation}
K^i_kK^k_j-K^i_jK^k_k+K\delta^i_j=0.\label{chareq}
\end{equation}
Using $g^{ij}$ to take the trace of Eq.~(\ref{chareq}) yields $K^i_jK^j_i=4H^2-2K$.
We expect that the term $\ m^iK_{ij}K^j_km^k$  leads to a preference for the tilt to order along the long axis of a prolate shape, and an increased bending stiffness along the direction parallel to $\mathbf{\hat m}$. The last term in the integrand of (\ref{threedtwod}) is like a spontaneous curvature given by a gradient of the projected director. Except for the fact that the anisotropic terms will lead to a preferred direction of $\mathbf{\hat m}$ relative to the principal directions of curvature, we expect that including these terms would not qualitatively change our results. A more dramatic change is expected if chiral terms like $m^i\epsilon_{ij}K^j_km^k$ are allowed. This term can give the vesicle a chiral shape, which is necessarily non-axisymmetric. It would be interesting to add this term alone to our minimal model and study vesicles shapes, as has recently been done for tubules and ribbons~\cite{TuSeifert2007}.

The most severe assumption of our model is the assumption of fixed defects at the poles. We have already mentioned that positive sign defects prefer regions of positive curvature, which suggests there may be lower energy, non-axisymmetric shapes than the ones we consider here. Furthermore, recent calculations have shown how a pair of defects of opposite sign can be pulled apart whenever there is a change in sign in the Gaussian curvature, with the positive defect migrating to the region of positive Gaussian curvature, and the negative defect migrating to the region of negative Gaussian curvature~\cite{VitelliNelson2004}. The magnitude of the curvature in both regions is important for determining whether or not a defect pair will unbind. Although we completely disregard this effect, we expect it will not play much of a role in the low-energy prolate shapes, since for those shapes there are at most only narrow bands of mildly negative Gaussian curvature where either end starts to bow out. An important extension of our calculation would be to lift the assumption of axisymmetry, and allow both the defect position and number and vesicle shape to vary. It would be interesting if the techniques of Ref.~\cite{ZhaoMahajanLuFang2005} could be used to experimentally determine defect position on vesicles with tilt order.

Finally, we have systematically studied the effect of changing the tilt modulus $K_m$ while deep in the ordered phase. While we have not presented results on the change in shape due to a phase transition from an ordered tilt phase to a disordered fluid phase, as in Ref.~\cite{MacKintoshLubensky1991}, we expect that the effect of increasing the order should be qualitatively similar to increasing $K_m$.

\begin{acknowledgments}
This work is supported in part by National Science Foundation
grant  NIRT-0404031 (TRP), and National Institutes of Health grant U54RR022232 (GH).  GH thanks the Richard Berlin Center for Cell Analysis \& Modeling for support. TRP thanks the Hatsopoulos Microfluids Laboratory at MIT, and TRP and GH thank the Aspen Center for Physics, where some of this work was done.
\end{acknowledgments}



\clearpage

\end{document}